\documentstyle[epsfig]{elsart}
\begin{document}
\begin{frontmatter}

\title{Incipient magnetic rotation? A magnetic dipole band in $^{104}$Cd}

\author[York]{D.~G. Jenkins\thanksref{Argonne}},
\author[York]{R. Wadsworth},
\author[McMaster]{J.~A. Cameron},
\author[ANL]{M.~P. Carpenter},
\author[SUNY]{C.~J. Chiara},
\author[LBNL]{R.~M. Clark},
\author[WashU]{M. Devlin},
\author[LBNL]{P. Fallon},
\author[SUNY]{D.~B. Fossan},
\author[York]{I.~M. Hibbert\thanksref{Liverpool}},
\author[ANL]{R.~V.~F. Janssens},
\author[AECL]{V.~P. Janzen},
\author[LBNL]{R. Kr\"ucken\thanksref{Yale}},
\author[SUNY]{D.~R. La~Fosse},
\author[SUNY]{G.~J. Lane\thanksref{LBL}},
\author[ANL]{T. Lauritsen},
\author[LBNL]{I.~Y. Lee},
\author[LBNL]{A.~O. Macchiavelli}, 
\author[York]{C.~M. Parry},
\author[WashU]{D.~G. Sarantites},
\author[SUNY]{J.~M. Sears},
\author[ANL]{D. Seweryniak},
\author[SUNY]{J.~F. Smith\thanksref{Manchester}},
\author[SUNY]{K. Starosta},
\author[LBNL]{D. Ward},
\author[ANL]{I. Wiedenhoever},
\author[York]{A.~N. Wilson},
\author[WashU]{J.~N. Wilson} and
\author[Rossendorf]{S. Frauendorf\thanksref{ND}}

\address[York]
{Department of Physics, University of York, Heslington, York Y01 5DD, U.K.}

\address[McMaster]
{Department of Physics and Astronomy, McMaster University, Hamilton, Ontario L85 4M1, Canada}

\address[LBNL]
{Lawrence Berkeley National Laboratory, Berkeley, CA 94720, USA}

\address[SUNY]
{Department of Physics, State University of New York 
at Stony Brook, Stony Brook, NY 11794-3800, USA}

\address[AECL]
{Chalk River Laboratories, AECL Research, Chalk River, 
Ontario K0J 1J0, Canada}

\address[WashU]
{Department of Chemistry, Washington University, St. Louis, MO 63130, USA}

\address[Rossendorf]
{FZ Rossendorf, Postfach 510119, D-01314 Dresden, Germany}

\address[ANL]
{Physics Division, Argonne National Laboratory, Argonne, IL 60439, USA}

\thanks[Argonne]
{ Present address: Physics Division, Argonne National Laboratory, Argonne, IL 60439, USA}

\thanks[Liverpool]
{ Present address: Oliver Lodge Laboratory, University of Liverpool, PO Box 147,  Liverpool L69 3BX, U.K.}

\thanks[Yale]
{ Present address: Wright Nuclear Structure Laboratory, Physics Department, Yale University, New Haven, CT 06520}

\thanks[LBL]
{ Present address: Lawrence Berkeley National Laboratory, Berkeley, California 94720}

\thanks[Manchester]
{ Present address: Nuclear Physics Group, Schuster Laboratory, University of Manchester, Brunswick Street, Manchester M13 9PL, U.K.}

\thanks[ND] { Present address: Nieuwland Science Hall, University of Notre Dame}

\begin{abstract}
High spin states of the nucleus $^{104}$Cd have been studied using the Gammasphere array. 
The level scheme for $^{104}$Cd has been revised and evidence for a structure consisting of magnetic dipole transitions is presented. 
Shell model calculations, published previously, are invoked to support an interpretation of this structure as an incipient case of magnetic rotation where the transversal magnetic dipole moment is not strong enough to break the signature symmetry.

\end{abstract}


\end{frontmatter}

\section{Introduction}

The shears mechanism (or magnetic rotation) is a mechanism whereby angular momentum is generated by the simultaneous alignment, from an intial perpendicular coupling, of long, rigid proton and neutron vectors with the total angular momentum vector \cite{tac}. 
The mechanism is characterised by a pronounced decrease in magnetic dipole transition strength, B(M1) as the shears vectors close. Such behaviour may be inferred from lifetime measurements.  
Using DSAM lifetime measurements, the existence of the shears mechanism was firstly confirmed in the light lead nuclei \cite{leads} and has more recently been shown to be manifested in the light tin nuclei, $^{106}$Sn and $^{108}$Sn \cite{snprl}. In the mass 110 region, the proton vector may be generated from g$_{9/2}$ holes in the Z=50 shell while the neutron vector is principally composed of h$_{11/2} $ neutrons and g$_{7/2}$,d$_{5/2}$ neutrons.
In the cadmium nuclei, 
recent lifetime measurements have confirmed the action of the shears mechanism  in $^{108}$Cd \cite{nsk}, $^{109}$Cd \cite{cd109} and $^{110}$Cd \cite{cd110}. 
However, the shears mechanism has not so far been shown to be active in the lighter cadmium isotopes.

Two of the outstanding issues associated with the phenomenon of magnetic rotation are the point of interface of both the shears mechanism with conventional collective rotation and the disappearance of the shears mechanism leading to the irregularly spaces shell model states.
The former issue is somewhat simpler in that rotation of the weakly deformed core may be modelled as a contributory vector to the total angular momentum which increases linearly with spin. 
However, it might be expected that the shears mechanism would be supplanted by collective rotation in nuclei for nuclei which have a deformation much larger than $\beta$$_{2}$ $\sim$ 0.1. Nevertheless, the exact point at which the transition between the magnetic rotation and collective regimes takes place is presently unclear.
Similarly, the lower boundary, so to speak, between the magnetic rotation regime and the spherical regime is uncertain. It is with this lower boundary that the present work is principlally concerned.

In terms of the cadmium nuclei,
the neutron number, N=56, forms an interesting dividing line since it separates the heavier cadmium nuclei whose high spin states are best understood in terms of collective rotational bands and the lighter cadmium nuclei which might be classified as spherical and whose limited valence space affords a relatively good understanding in terms of the shell model.
The nucleus $^{104}$Cd might therefore be a good testing ground for an investigation of  
the boundary between the spherical nuclei and the region where magnetic rotation is possible.

\section{Experimental Details and Data Analysis}

Excited states in the nucleus $^{104}$Cd were populated using the $^{54}$Fe($^{58}$Ni,$\alpha$4p) reaction at a beam energy of 243 MeV in two separate experiments. In the first experiment, a $^{58}$Ni beam accelerated by the 88-inch cyclotron at the Lawrence Berkeley National Laboratory was incident on a target composed of 600$\mu$g/cm$^{2}$ enriched $^{54}$Fe on a backing of 15.2mg/cm$^{2}$ of gold.
The full implementation of the Gammasphere array \cite{gammasphere} with
95 HPGe detectors was used to detect the resulting $\gamma$ decay. In a second experiment a $^{58}$Ni beam from the ATLAS accelerator at Argonne National Laboratory was directed onto a 540$\mu$g/cm$^{2}$ enriched $^{54}$Fe thin target. The $\gamma$ decay was studied using the Gammasphere array consisting of 100 HPGe detectors. The array was coupled with the Microball charged particle detector system \cite{microball} in order to provide channel selection. A large number of nuclei in the mass 100 to 110 region were populated in this reaction; the strongest channels being $^{109}$Sb and $^{108}$Sn with 27\% and 42\% of the total cross-section, respectively. However, $\gamma$ rays from the decay of $^{104}$Cd, 
the subject of this study, comprised only approximately 6.5\% of the data-set. The next lightest even-even cadmium nucleus, $^{102}$Cd, whose high-spin structure has already been thoroughly studied \cite{cd102}, was populated with a similar intensity ($\approx$6\%).

\begin{figure} [h]
\centerline{\epsfig{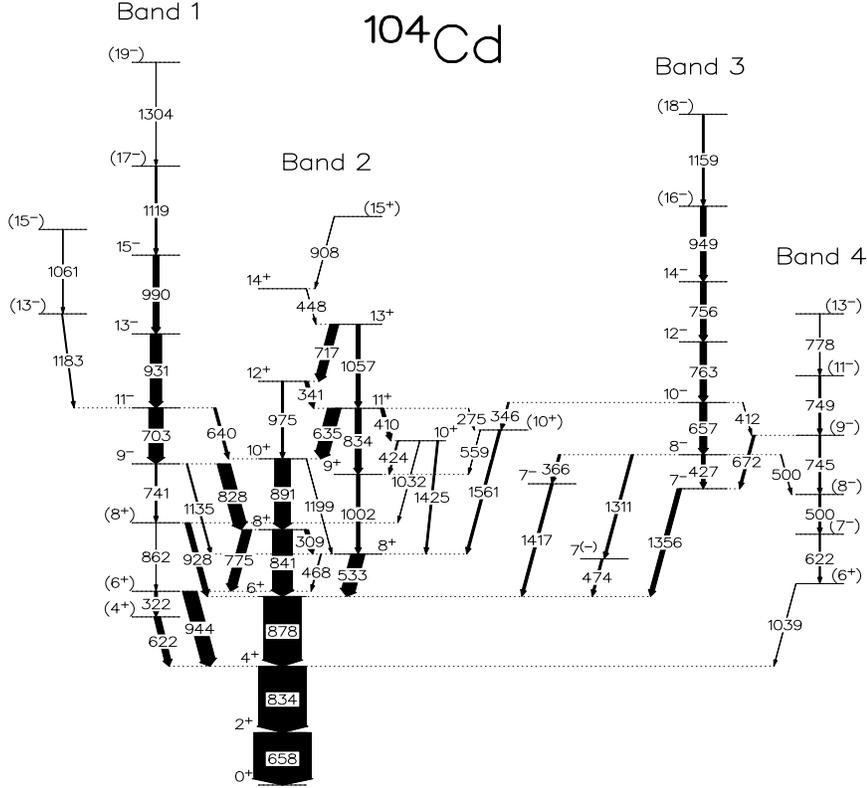}}
\caption{Partial decay scheme for $^{104}$Cd. The width of the arrows relates to the measured intensity of the transitions.}
\label{fig1}
\end{figure}

The backed-target data were unfolded and used to produce a $\gamma$-$\gamma$-$\gamma$ cube, containing 2.1$\times$10$^{10}$ triples events, which was analysed 
using the RADWARE analysis program, Levit8r \cite{radware}.
The thin-target data were unfolded and a channel selection requirement of one $\alpha$ particle and at least 2 protons was made. This allowed a $\gamma$-$\gamma$-$\gamma$ cube to be created, containing 9.2$\times$10$^{8}$ triples, which had reasonable selection for the $\alpha$4p channel without overly reducing the statistics. The absence of the hevimet collimators on the BGO shields for the thin target experiment permitted total energy (H) and $\gamma$-ray multiplicity (K) to be measured. Accordingly, suitable two-dimensional coincidence gates were set on H and K in order to enhance the desired reaction channel. The gates were set by identifying the region in the H-K plane associated with the $\gamma$ rays at the bottom of the $^{104}$Cd level scheme. Both the thick- and thin-target cubes were explored for detailed information about the structure of $^{104}$Cd ; the resulting level scheme is presented in Figure~\ref{fig1}. The level scheme has been considerably revised from published level schemes for the low-spin shell model states \cite{cd104isomers,cd104old} and high-spin rotational bands
\cite{klamra}.
During the preparation of this work, a work was published which revised the prior level schemes and discussed some aspects of the revised scheme in terms of the interacting boson-fermion model (IBM/IBFM) \cite{deang}. The level scheme presented in the present work is in broad agreement with this recent publication.

In order to perform an angular correlation analysis, the backed target data were unfolded and sorted into two 2D matrices. In each case the data were unpacked into triples and gates were set on known E2 transitions (658.0, 834.3, 878.4, 840.9 and 890.5 keV) at the bottom of the $^{104}$Cd level scheme in order to enhance the $\alpha$4p reaction channel.  Two angle sets were used for the angular correlation analysis. Angle Set 1 comprised those detectors at far forward and backward positions including detectors at 17$^{\circ}$, 32$^{\circ}$ and 37$^{\circ}$ and the corresponding detectors in the other hemisphere, at 143$^{\circ}$, 148$^{\circ}$ and 163 $^{\circ}$. Angle Set 2 comprised those detectors close to 90$^{\circ}$  including detectors at 79$^{\circ}$, 81$^{\circ}$, 90$^{\circ}$, 99$^{\circ}$ and 101$^{\circ}$. These particular groups were chosen since they contained a similar number of detectors and were large enough to make good use of the angular composition of the array while clearly discriminating between dipoles and quadrupoles. Matrices were created for Set 1 against Set 1 and Set 1 against Set 2.  An angular correlation method \cite{DCO} was used to assign the multipolarity of previously unknown transitions, using the first two matrices. These matrices were used to extract angular correlation ratios (R$_{DCO}$) from the ratio of the intensities of transitions in angle set 1 to those in angle set 2, after gating on known stretched quadrupoles in angle set 1. The values were normalized to take into account the difference between the numbers of detectors in angle set 1 (30) and those in angle set 2 (37). Angular correlation ratios for this geometry were obtained for transitions of known multipolarity in $^{108}$Sn as a guide in assigning multipolarites of unknown transitions in $^{104}$Cd. By this means, it was deduced that stretched dipole - stretched quadrupole transitions from this geometry have an average R$_{DCO}$ value of 1.45 whilst those of known stretched quadrupole - stretched quadrupole transitions have an average R$_{DCO}$ value of 1.85. The results of this angular correlation analysis are
presented in Table 1.
In addition to the angular correlation matrices, a further matrix was sorted containing all detectors against all detectors and this was used to obtain intensity information and hence to extract branching ratios.  

\begin{figure}[h]
\centerline{\epsfig{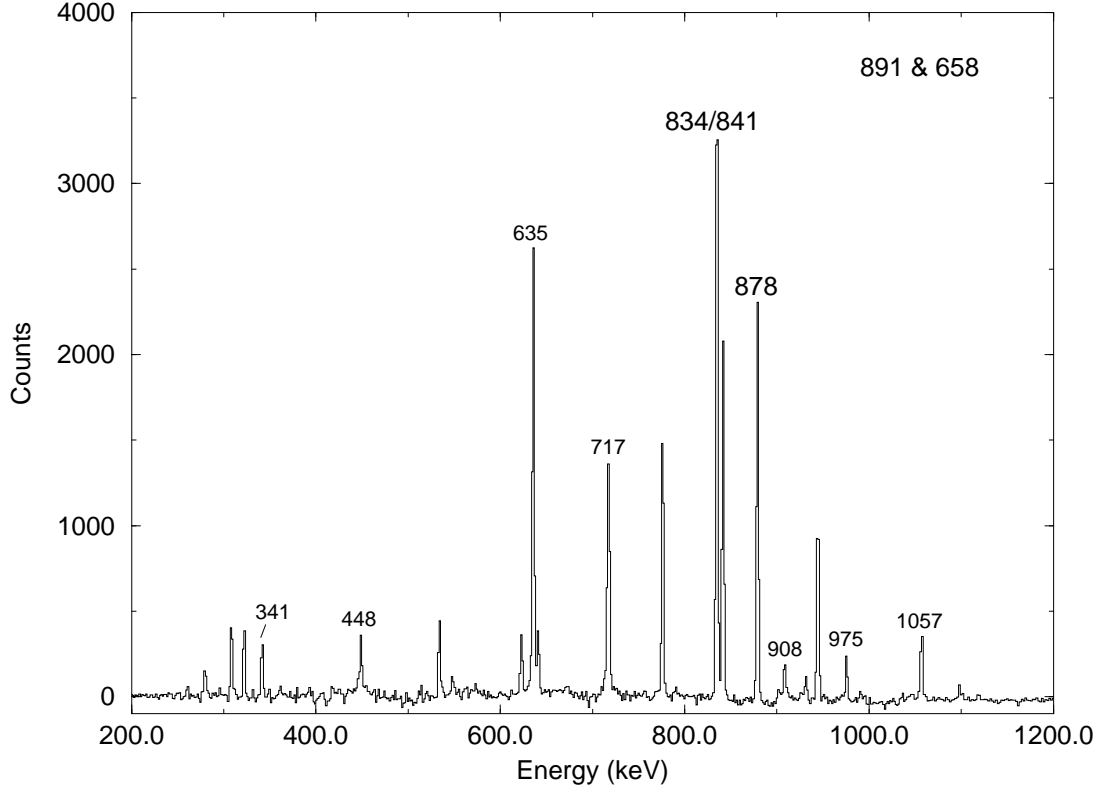}}
\caption{Spectrum of band 2, double-gated on the 891 and 658 keV transitions in the thick-target cube with the band members 
indicated in the smaller type.}
\label{band2}
\end{figure}

Band 2 was originally classified by Klamra {\em et al.} \cite{klamra} as a decoupled band, although the irregularity of the observed transitions and the sharply contrasting features of this band with similar structures in neighbouring isotopes were noted.
In common with the recent work of De Angelis {\em et al.} \cite{deang}, this structure is interpreted in the present work as a cascade of dipole transitions, crossed-over by at least two E2 transitions (figure~\ref{band2}). The multipolarity of the dipole transitions has been verified by angular correlation measurements up to the 13$^{+}$ level. Due to the combination of the low statistics in this weakly populated channel and the design of the Gammasphere array, it was not possible to measure the linear polarisation of these dipole transitions to clearly establish whether they are of electric or magnetic character. However, if the dipoles are assumed to be E1 then the intensity of the dipole transitions would imply B(E1)/B(E2) ratios which are an
order of magnitude larger than those of the deformed octupole nuclei in the mass 220 region and several orders of magnitude larger than octupole vibrational effects observed in the A$\sim$110 region, e.g. $^{108}$Te \cite{te108}. It is therefore assumed that these transitions are of M1 multipolarity. 

It should be noted that the structure of band 2 presented here differs from that presented in ref \cite{deang} in that the ordering of the 635 and 717 keV transitions is reversed. 
The alternate ordering given in ref \cite{deang} is made on the basis of the existence of a structure parallel to band 2 which contains a 636 keV $\gamma$-ray. The present data-set does not support the existence of such a structure or a 635/636 keV doublet. Furthermore, coincidence data confirm the ordering suggested here.

\section{Discussion}

Since the recent work of de Angelis {\em et al.} \cite{deang} provides a comprehensive description in terms of the IBM/IBFM model of the decoupled bands observed in $^{104}$Cd, it is not the intention of the present work to rehearse the configuration assignments appropriate to bands 1,3 and 4. However, this recent work \cite{deang} did not consider the possibility of two g$_{9/2}$ proton holes coupled to spin 8$^{+}$ within the IBM model. This approach was, therefore, unable to account for the properties of band 2. Accordingly, the remainder of this discussion is focussed on the reasons for the appearance of the cascade of magnetic dipole transitions, band 2.

It is instructive to look to the neighbouring even-even cadmium nuclei, $^{102}$Cd and $^{106}$Cd. $^{106}$Cd does not exhibit a dipole structure analagous to that observed in $^{104}$Cd. Instead, there is a decoupled band built on the 10$^{+}$ bandhead which is believed to be generated from an aligned $\nu$(h$_{11/2}$)$^{2}$ configuration \cite{cd106}. De Angelis {\em et al.} suggests that a decoupled band with a 12$^{+}$ bandhead, not observed in the present work, is the analogy of this structure in $^{104}$Cd \cite{deang}.
However, a structure similar to the dipole structure in $^{104}$Cd has been observed in $^{102}$Cd \cite{cd102}. In the case of the $^{102}$Cd band, it was tentatively suggested
that some kind of shears mechanism might be responsible for the appearance of the irregular dipole band in this nucleus \cite{cd102}. The two g$_{9/2}$ proton holes in the Z=50 shell could couple to a spin of 8$^{+}$ providing the proton vector of the shears, whilst the neutron vector might be composed of d$_{5/2}$,g$_{7/2}$ neutrons coupled to 6$^{+}$ \cite{cd102}. Such a shears configuration could produce a shears band with a 10$^{+}$ bandhead.
However, one of the characteristic features of the known shears bands is their regular increase in excitation energy with increasing spin e.g. \cite{dgj}. Clearly, the dipole band in $^{102}$Cd is highly irregular and does not exhibit such a smooth increase. Therefore, this interpretation of the $^{102}$Cd band does not fit easily into the conventional picture of a shears band. Although similar particles are active in the dipole bands in $^{102,104}$Cd as in the shears bands in $^{108-110}$Cd, it is clear that they do not both result in shears bands. 

\begin{figure}[!h]
\centerline{\epsfig{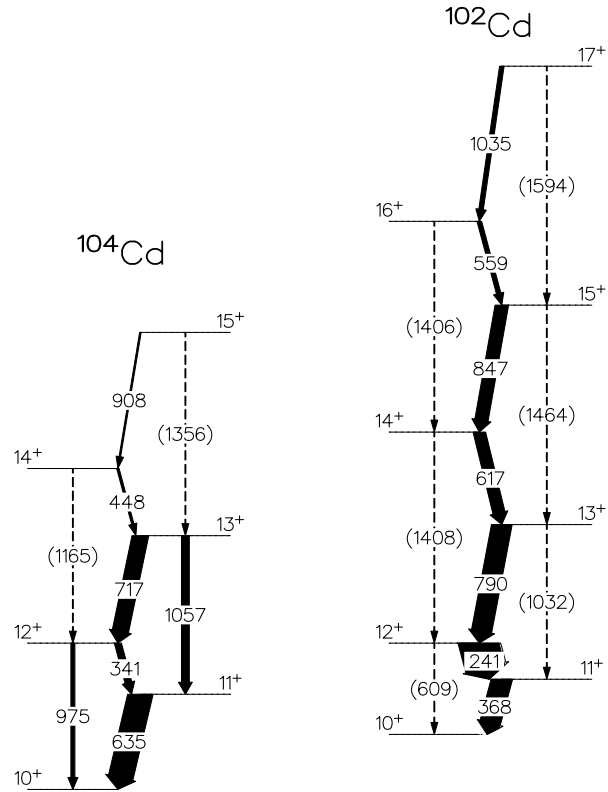}}
\caption{Partial decay schemes of the analogous structures in $^{102}$Cd (taken from ref \cite{cd102}) and band 2 in $^{104}$Cd. The 12$^{+}$ levels have been
plotted parallel to each other to aid the comparison.}
\label{comparison}
\end{figure}

A comparison of the dipole structures in $^{102}$Cd and $^{104}$Cd in figure~\ref{comparison} makes it clear that neither band exhibits the regular increase in transition energy observed in the shears bands in the heavier cadmium isotopes.
However, they do share common features in that they both have 10$^{+}$ bandheads and the odd spin to even spin dipole transitions have much larger energies than their even spin to odd spin counterparts. The similarity of the two structures points to a common underlying structure, most probably in the coupling of the g$_{9/2}$ holes to the d$_{5/2}$,g$_{7/2}$ neutrons. 

As a means of understanding these irregular structures in the light cadmium nuclei, it is instructive to refer to previous work on magnetic dipole structures in the light indium nuclei, which highlights the question of regularity in magnetic dipole structures \cite{india}. $^{107}$In exhibits an irregular sequence of M1 transitions with fluctuating transition energies, whereas the heavier nuclei, $^{109,111,113}$In have structures of 5-6 transitions that have similar characteristics to shears bands and become more regular with increasing mass. It has been suggested that a kind of shears mechanism is taking place in these nuclei between a pair of h$_{11/2}$ neutrons and a g$_{9/2}$ proton hole. Depending
on the position in the shell, the d$_{5/2}$,g$_{7/2}$ neutrons, which are spectating, may 
couple to the shears mechanism taking place, augmenting it in some manner, or they may disrupt it leading
to an irregular sequence of dipoles. Frauendorf and Reif distinguish between four scenarios for generating angular momentum using the g$_{7/2}$,d$_{5/2}$ particles \cite{india}. These scenarios depend on the particular coupling and quadrupole moments of the g$_{7/2}$,d$_{5/2}$ neutrons. 

\begin{itemize}

\item The quadrupole moment of the g$_{7/2}$,d$_{5/2}$ neutrons is negative.  Spin is generated by the recoupling of the g$_{7/2}$,d$_{5/2}$ neutrons, leading to the irregular multiplets seen in near-spherical nuclei.

\item The quadrupole moment is positive. The g$_{7/2}$,d$_{5/2}$ neutrons orient perpendicular to the $\nu$[h$_{11/2}$$^{2}$]$_{J=10}$ pair, generating angular momentum from a shears mechanism between these two ``blades'' adding to that shears mechanism occurring between the g$_{9/2}$ proton hole and the  $\nu$[h$_{11/2}$$^{2}$]$_{J=10}$ pair.

\item The quadrupole moment of the g$_{7/2}$,d$_{5/2}$ neutrons is close to zero. The g$_{7/2}$,d$_{5/2}$ neutrons orient themselves parallel to the $\nu$[h$_{11/2}$$^{2}$]$_{J=10}$ pair. This augments the shears mechanism taking place, again, leading to a regular structure. 

\item The g$_{7/2}$,d$_{5/2}$ neutrons couple to low spin (J=0 or 2). The J=2 admixture slowly increases generating spin from the collective rotation of the slightly deformed $\nu$[g$_{7/2}$,d$_{5/2}$] system.

\end{itemize}

 There are four g$_{7/2}$,d$_{5/2}$ neutrons above the Z=50 shell in $^{102}$Cd and
six g$_{7/2}$,d$_{5/2}$ neutrons in $^{104}$Cd. Shell model calculations using RITSSCHIL \cite{ritsschil} give Q $\approx$ -33 e fm$^{2}$ for
(g$_{7/2}$,d$_{5/2}$)$^{4}$$_{J=6}$ and Q $\approx$ -7 e fm$^{2}$ for (g$_{7/2}$,d$_{5/2}$)$^{6}$$_{J=6}$ \cite{india}. One is tempted, therefore, to suggest that the negative quadrupole moment of $^{102}$Cd leads to the irregular structure observed in $^{102}$Cd which arises from the first mechanism, while the less negative quadrupole moment for $^{104}$Cd leads to a somewhat more ``regular'', albeit staggered, structure. However, the quadrupole moment is not sufficiently near to zero for the structure to go over into the shears mechanism via the the third mechanism described above.

Considering the fact that the g$_{9/2}$ holes are coupled to 8$^{+}$ and have a large positive g-factor and the 
g$_{7/2}$,d$_{5/2}$ neutrons involved a low g factor, the effective g-factor for the proposed coupling will be large. In turn, this implies relatively large absolute B(M1) values and short lifetimes for these states. However, the odd-even staggering of the states in band 2 implies that the transversal magnetic moment does not break the signature symmetry to any great extent. Whilst it is clear from the backed-target data that the dipole transitions in this band are considerably Doppler-shifted, there is, unfortunately, insufficient intensity to perform a lifetime analysis.
 
\section{Conclusion}

In summary, a cascade of magnetic dipole transitions, band 2, has been observed in $^{104}$Cd and interpreted as a case of incipient magnetic rotation.
This dipole band is very similar to a structure in $^{102}$Cd but has a somewhat more
regular appearance. Shell model calculations indicate that the coupling of the
g$_{7/2}$,d$_{5/2}$ neutrons is highly sensitive to their quadrupole moments. Furthermore, it has been shown from previous work that this can result in regular or irregular magnetic dipole bands depending on whether these neutrons have a positive or negative quadrupole moment. The increase in the number of positive parity neutrons from $^{102}$Cd to $^{104}$Cd may explain the origin of the increased regularity of the dipole band in $^{104}$Cd. The odd-even spin state staggering observed suggests that the symmetry breaking is moderate and that it is more appropriate to describe this structure in terms of the spherical shell model rather than in terms of magnetic rotation.
This work provides useful information in determining the boundaries of and conditions for the appearance of magnetic rotation in the mass 110 region.

\ack {We would like to thank the staff of both the 88--Inch Cyclotron and the ATLAS accelerator and A. Lipski for making the targets used. This work has been supported in part by the U.S. NSF and Department of Energy under Contracts No. DE--AC03--76SF00098 and W-31-109-ENG-38, the UK EPSRC, AECL and the Canadian NSERC. RW and RMC acknowledge receipt of a NATO grant.}

\begin{table}
\caption{Spectroscopic information for $^{104}$Cd from the present work including $\gamma$ ray energies, intensities relative to the 834.3 keV and DCO
ratios obtained by the method described in the text.\label{Table1}}
\begin{center}
\begin{tabular}{ccccccc}
\hline
\hline
\\
E$_\gamma$(keV) &I$_\gamma$(\%) &R$_{DCO}$ &M$_{\lambda}$ &Assignment\\ 
\\
\hline
274.8(4) & 0.7(1)    &           &        &$11^+\rightarrow(10^+)$\\
308.6(2) & 8.7(2)    & 1.17(09)  &  M1/E2 &$8^+\rightarrow8^+$\\
321.6(2) & 6.0(2)    & 1.16(13)  &  E2    &$(6^+)\rightarrow(4^+)$\\
341.2(2) & 7.5(2)    & 1.28(10)  &  M1    &$12^+\rightarrow11^+$\\
346.0(3) & 3.2(2)    &           &        &$10^-\rightarrow(10^+)$\\
365.7(3) & 4.7(1)    & 0.63(19)  &  M1/E2 &$8^-\rightarrow7^-$\\
409.8(2) & 7.2(3)    & 1.43(15)  &  M1    &$11^+\rightarrow10^+$\\
411.8(4) & $<$0.5    &           &        &$10^-\rightarrow9^-$\\
423.8(3) & 3.6(1)    & 1.45(32)  &  M1    &$10^+\rightarrow9^+$\\
427.2(2) & 7.7(2)    & 0.81(15)  &  M1/E2 &$8^-\rightarrow7^-$\\
448.0(3) & 2.6(3)    & 1.31(26)  &  M1    &$14^+\rightarrow13^+$\\
467.6(3) & 1.9(1)    & 	         &        &$8^+\rightarrow6^+$&\\
473.7(2) & 4.4(2)    &           &        &$(7^-)\rightarrow6^+$&\\
499.8(3) & 5.6(3)$^{\dagger}$ &           &        &$(8^-)\rightarrow(7^-)$&\\
499.8(3) & 5.6(3)$^{\dagger}$ &           &        &$(8^-)\rightarrow(8^-)$&\\
533.3(1) & 28.5(3)   & 1.89(09)  &  E2    &$8^+\rightarrow6^+$&\\
559.1(3) & 3.0(3)    &           &        &$(10^+)\rightarrow9^+$&\\
622.1(1) & 13.3(2)$^{\dagger}$   &           &        &$(4^+)\rightarrow4^+$\\
622.1(2) & 13.3(2)$^{\dagger}$   &           &        &$(7^-)\rightarrow(6^+)$&\\
635.4(1) & 27.5(3)   & 1.41(06)  &  M1    &$11^+\rightarrow10^+$&\\
640.4(1) & 3.5(2)    &           &        &$11^-\rightarrow10^+$&\\
657.0(1) & 14.7(3)   & 1.84(13)  &  E2    &$(11^-)\rightarrow(9^-)$&\\
658.0(1) & -         &           &  E2    &$2^+\rightarrow0^+$&\\
671.7(3) & 5.2(3)    &           &        &$(9^-)\rightarrow7^-$&\\
703.1(1) & 30.0(3)   & 1.86(04)  &  E2    &$11^-\rightarrow9^-$&\\

\\
\hline
\end{tabular}
\end{center}
\end{table}

\newpage

\begin{table} [t]
\begin{center}
\begin{tabular} {ccccccc}
\hline
\hline
\\
E$_\gamma$(keV) &I$_\gamma$(\%) &R$_{DCO}$ &M$_{\lambda}$ &Assignment\\ 
                \\
\hline

716.9(1) & 18.2(3)   & 1.44(07)  &  M1    &$13^+\rightarrow12^+$&\\
740.8(3) & 3.3(1)    &  	 &        &$9^-\rightarrow(8^+)$&\\
744.6(3) & 3.4(3)    &           &        &$9^-\rightarrow(8^-)$&\\
748.8(3) & 4.0(1)    &           &        &$(11^-)\rightarrow(9^-)$&\\
755.7(2) & 11.9(2)   & 1.85(14)  &  E2    &$14^-\rightarrow12^-$&\\
762.9(1) & 13.6(2)   & 1.92(19)  &  E2    &$12^-\rightarrow10^-$&\\
775.3(1) & 19.4(2)   & 1.86(11)  &  E2    &$8^+\rightarrow(6^+)$&\\
778.0(3) &  1.3(1)   &           &        &$(13^-)\rightarrow(11^-)$&\\
827.7(1) & 26.9(2)   & 1.49(06)  &  E1    &$9^-\rightarrow8^+$&\\
833.5(2) & 12.7(8)   &           &        &$11^+\rightarrow9^+$&\\
834.3(1) &$\equiv$ 100&          &  E2    &$4^+\rightarrow2^+$&\\
840.9(1) & 41.8(2)   & 1.82(05)  &  E2    &$8^+\rightarrow6^+$& \\
862.2(3) & 2.6(3)    &           &        &$(8^+)\rightarrow(6^+)$&\\
878.4(1) & 78.5(3)   & 1.85(04)  &  E2    &$6^+\rightarrow4^+$& \\
890.5(1) & 32.9(2)   & 1.86(12)  &  E2    &$10^+\rightarrow8^+$& \\
907.7(3) & 1.7(2)    &           &        &$(15^+)\rightarrow(14^+)$&\\
927.9(2) & 10.9(2)   & 1.94(18)  &  E2    &$(8^+)\rightarrow6^+$&\\
931.0(1) & 23.6(2)   & 1.86(06)  &  E2    &$13^-\rightarrow11^-$&\\
943.9(1) & 29.7(2)   & 1.54(12)  &        &$(6^+)\rightarrow4^+$&\\
949.0(2) & 11.5(2)   &           &        &$16^-\rightarrow14^-$&\\
974.5(2) & 4.4(2)    & 2.01(22)  &  E2    &$12^+\rightarrow10^+$&\\
989.7(2) & 12.5(3)   & 1.87(13)  &  E2    &$15^-\rightarrow13^-$&\\
1001.6(2)& 6.9(2)    & 1.34(13)  &        &$9^+\rightarrow8^+$&\\
1031.6(3)& 1.9(1)    &           &        &$10^+\rightarrow(8^+)$&\\
1039.4(3)& 2.0(2)    &           &        &$(6^+)\rightarrow4^+$&\\
1056.6(2)& 8.7(2)    & 1.88(21)  &  E2    &$13^+\rightarrow11^+$&\\
\\
\hline
\end{tabular}
\end{center}
\end{table}

\newpage

\begin{table} [t]
\begin{center}
\begin{tabular} {ccccccc}
\hline
\hline
\\

E$_\gamma$(keV) &I$_\gamma$(\%) &R$_{DCO}$ &M$_{\lambda}$ &Assignment\\ 
                \\
\hline

1061.2(4)& 1.8(3)    &           &        &$(15^-)\rightarrow(13^-)$&\\
1119.4(3)& 3.7(3)    &           &        &$(17^-)\rightarrow15^-$&\\
1135.4(3)& 2.7(3)    &           &        &$9^-\rightarrow8^+$&\\
1159.2(4)& 3.8(3)    &  	 &        &$(18^-)\rightarrow(16^-)$&\\
1183.2(4)& 2.0(2)    &           &        &$(13^-)\rightarrow11^-$&\\
1199.2(4)& 1.0(2)    &           &        &$(10^+)\rightarrow8^+$&\\
1304.2(4)& 0.6(2)    &           &        &$(19^-)\rightarrow(17^-)$&\\
1311.3(6)& 1.1(3)    &           &        &$8^-\rightarrow(7^-)$&\\
1355.8(2)& 9.5(2)    & 1.35(18)  &  E1    &$7^-\rightarrow6^+$&\\
1417.4(3)& 5.6(2)    & 1.52(23)  &  E1    &$7^-\rightarrow6^+$&\\
1425.2(3)& 3.6(1)    &           &        &$(10^+)\rightarrow8^+$&\\
1561.1(5)& 0.8(1)    &           &        &$(10^+)\rightarrow8^+$&\\

\end{tabular}
\end{center}
\protect$^{\dagger}$ Intensity given for doublet\protect\\
\end{table}

\pagebreak

\end{document}